\documentclass[journal,comsoc]{IEEEtran}
\usepackage{amsmath,amssymb,amsfonts,graphicx,mathtools,nccmath,amsthm,float}
\usepackage{multicol,tikz,cite,array,booktabs,url}
\usepackage{enumitem}
\usepackage{cite}
\usepackage{textcomp}
\usepackage{xcolor}
\usepackage{algorithm}
\usepackage{algpseudocode,algorithmicx}
\usepackage{acronym}
\usepackage[bookmarks,colorlinks]{hyperref}

\begin{document}

\title{Lightweight Security for\\ Ambient-Powered Programmable Reflections with\\ Reconfigurable Intelligent Surfaces}
\author{Andreas~Kunz,~\IEEEmembership{Senior~Member,~IEEE,} Sheeba Backia Mary Baskaran, \\and George~C.~Alexandropoulos,~\IEEEmembership{Senior~Member,~IEEE}
\thanks{A. Kunz and S. B. M. Baskaran are with the Lenovo/Motorola Mobility, 61440 Oberursel, Germany (e-mails: \{akunz, smary\}@lenovo.com).}
\thanks{G. C. Alexandropoulos is with the Department of Informatics and Telecommunications, National and Kapodistrian University of Athens, 15784 Athens, Greece and with the Department of Electrical and Computer Engineering, University of Illinois Chicago, Chicago, IL 60601, USA (e-mail: alexandg@di.uoa.gr).}
} 

\maketitle

\begin{abstract}
Ambient Internet-of-Things (AIoT) form a new class of emerging technology that promises to deliver pervasive wireless connectivity to previously disconnected devices and products, assisting dependent industries (for example, supply chain, clothing, remote surveillance, climate monitoring, and sensors) to obtain granular real-time service visibility. Such ultra-low complexity and power consumption devices, that are either battery-less or have the capability for limited energy storage, can provide data feeds about the condition of any aspect (e.g., an environment or an item) that is being monitored, enabling proactive or reactive control by any application server. Although the security of data involving AIoT devices is critical for key decisions of any dependent operational system, the implementation of resource intensive cryptographic algorithms and other security mechanisms becomes nearly infeasible, or very challenging, due to the device energy and computational limitations. In this article, we present a lightweight security solution that enables confidentiality, integrity, and privacy protection in wireless links including AIoT. We consider, as a case study, an ambient-powered Reconfigurable Intelligent Surface (RIS) that harvests energy from its incident radio waves to realize programmable reflective beamforming, enabling the communication between a Base Station (BS) and end-user terminals. The proposed lightweight security solution is applied to the control channel between the BS and the RIS controller which is responsible for the metasurface's dynamic management and phase configuration optimization. 
\end{abstract}

\begin{IEEEkeywords}
Ambient energy, 5G-Advanced, 6G, 3GPP Release $19$, security, IoT, reconfigurable intelligent surface.
\end{IEEEkeywords}

\section{Introduction} \label{Sec:Intro}
Internet-of-Things (IoT) empowered by ambient sources is currently being specified in $3$rd Generation Partnership Project ($3$GPP) Release $19$~\cite{3gppTR2370013} (i.e., 5G-Advanced), where new service requirements are identified along with new key performance indicators. Ambient IoT devices are being either battery-less or come with limited energy storage capability (i.e., using a capacitor), where energy is provided through the harvesting of radio waves, light, motion, heat, or any other power source that can be seen suitable~\cite{Ambient_IoT_mag2022}. The latest requirements cover low complexity, zero maintenance, long life span (e.g., more than $10$ years), and small size, as well as lower capabilities and power consumption than previously defined $3$GPP IoT devices (e.g., Narrowband IoT (NB-IoT) and enhanced Machine Type Communication (eMTC) devices). Based on the currently envisioned business models, it is anticipated that the majority of ambient IoT devices will not have a Universal Subscriber Identity Module (USIM), as a normal $3$GPP-defined User Equipment (UE) possesses, although, they will require certain level of security for wireless communications. 

Several use cases and requirements for ambient IoT require not only confidentiality protection of communications, but also their integrity protection. However, these two forms of security protection, as realized by the relevant available $3$GPP algorithms described in \cite{3gppTS33501}, require a USIM and a respective memory, as well as processing power. Especially for integrity protection, this would not be feasible for the envisioned ambient IoT use cases without a battery and harvested energy. Hence, it is still an open challenge to devise a low-power computational method for acceptable confidentiality and integrity protection levels for communications, considering the minimum possible number of transmitted/received messages and processing limitations of Ambient IoT devices~\cite{IoT_AmI_2022}. In addition, further privacy protection is not assured if the same identity is used, thus, the communication could be easily tracked by any eavesdropper(s), and further attacks could be carried out towards the ambient IoT device and the server.

The security group within $3$GPP has performed several security studies for interconnecting IoT devices and Machine-to-Machine (M2M) communications. All those studies are focusing on different security aspects, still considering, however, a USIM as the main basis for implementing security, including confidentiality and integrity protection~\cite{3gppTS33501}. This requirement can be relevant for M2M/IoT devices fixedly installed somewhere for a long-lasting operation (e.g., a sensor under a bridge to measure the water level), where the size, including the battery size, does not matter much. Other IoT devices in industrial environments, as described in the last 3GPP discussion, can be also fixedly installed having a fixed power supply, and the focus has been more to date on ultra-reliable and low-latency communications to transfer information between machines for smooth production processes. 

One example of an emerging IoT device is a Reconfigurable Intelligent Surface (RIS)~\cite{etsiGR_RIS003}, which comprises multiple ultra-thin unit elements each with multiple digitalized states corresponding to distinct electromagnetic responses. The phase configuration of these elements, which can take place ultra fast and with ultra-low power consumption (power is needed for the response tunability and maintenance), is controlled by a very low power controller that is also tasked to interface the overall RIS device with the rest of the network~\cite{RIS_Stc_mag}. RISs are intended to coat building facades and indoor walls (static devices) or vehicles (nomadic devices) extending the coverage of Base Stations (BSs, or gNB in 5G terminology)~\cite{EURASIP_RIS_all}, and can be optimized to manipulate information-bearing signal propagation directly in the wave domain to ``trusted zones'' (such as home, business space, residential area, and restricted service area), reducing data leakage to potential eavesdroppers and enhancing communication security~\cite{RIS_counteracting_2023}. To this end, RISs can be designed to avoid the leakage of confidential signals to eavesdroppers via specular reflections, playing a significant role in improving the communication security of other ambient IoT devices deployed in certain trusted areas. 

The aforedescribed, widely considered in the literature and in recent demonstrations, very low power RIS devices, that are deprived of active transmit/receive components, usually include a simple Printed Circuit Board (PCB) controller, to whom is being assigned to load the optimized phase configurations for all unit elements of the metasurface~\cite{RISoverview2023_all}. Those optimized element responses are usually computed on an external device, usually the BS, that wishes to utilize and/or owns the RIS with the objective to realize almost energy-neutral coverage extension. This mechanism requires an in-band or an out-of-band control communication channel between the RIS controller and the BS~\cite{RISE6G_COMMAG}, which is independent of the RIS-assisted relaying mechanism of the radio signals from/to the BS to/from one or more User Equipment (UE). 

In this article, we present a lightweight security solution enabling confidentiality, integrity, and privacy protection for the control communication channel between an ambient-powered RIS and the BS that utilizes it. The proposed procedure is based on the recent relevant proposals in the technical report~\cite{3gppTR33713}, and is valid for securing communication links including any type of an ambient IoT device.

\section{Ambient-Powered RIS Devices} \label{Sec:RIS}
\begin{figure}[t!]
\centering
\includegraphics[width=\columnwidth]{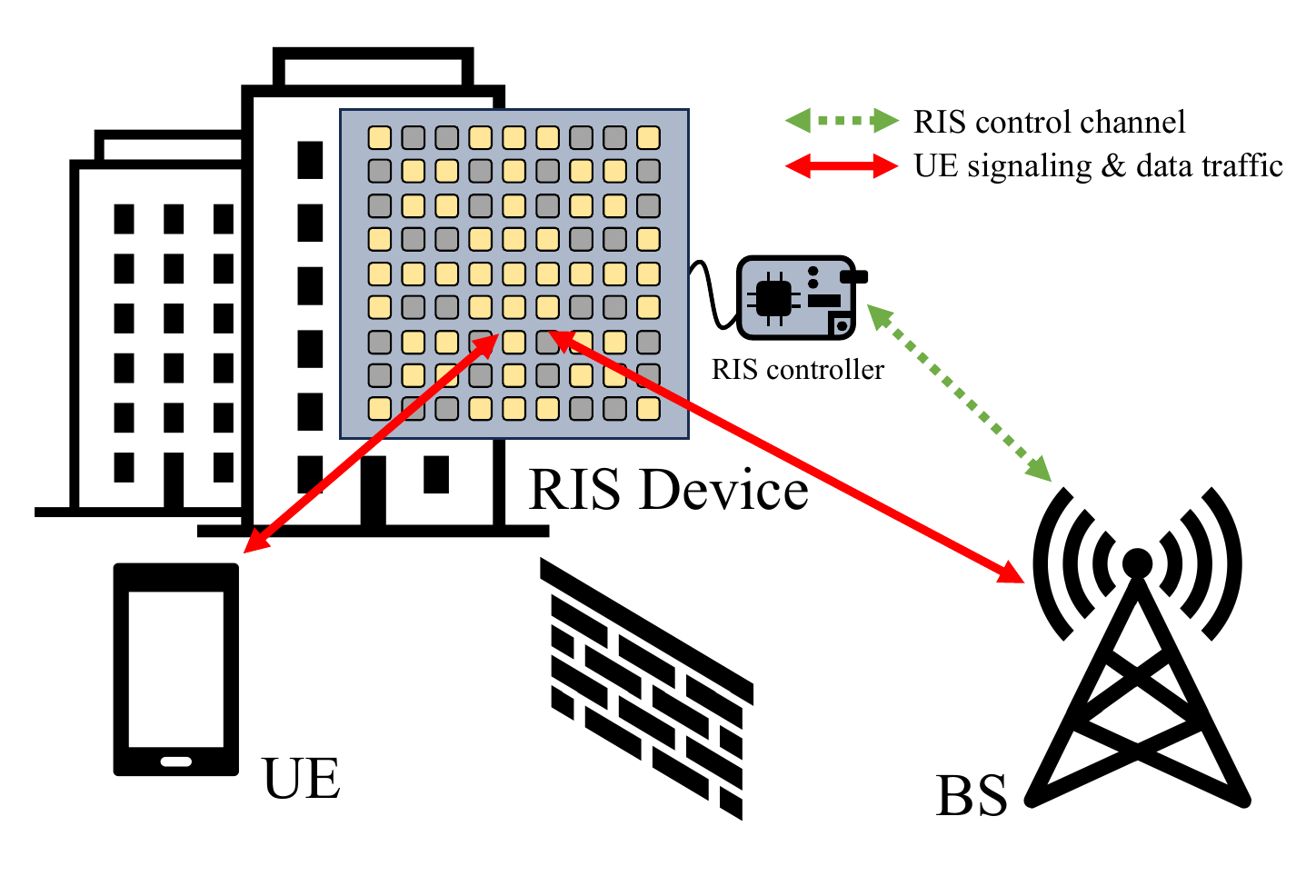}
\caption{Wireless communications between a Base Station (BS) and a User Equipment (UE) via an ambient-powered Reconfigurable Intelligent Surface (RIS). The RIS is controlled by the BS which shares the optimum phase configuration with the RIS controller via a dedicated secure control channel.}
\label{fig:RIS_control_channel} 
\end{figure}
A typical RIS-empowered wireless communication system between one BS and one UE is illustrated in Fig.~\ref{fig:RIS_control_channel}. We consider an very low power RIS device comprising unit elements, of tunable responses to impinging electromagnetic waves (the RIS panel), and a PCB controller (the RIS controller). The latter is missioned to retain and/or change the state of each of those elements as well as to exchange control communication signals with one or both the end devices. In this article, without loss of generality, we assume that the BS initiates and maintains a bidirectional in-band or out-of-band control channel with the RIS controller, i.e., the overall RIS device, and our intention is to design confidentiality, integrity, and privacy protection mechanisms for this channel. Following~\cite{Saggese_RIScontrol}, the RIS control channel can be used to trigger the RIS controller to sweep among predefined RIS phase configuration profiles to either enable searching for the one yielding acceptable end-to-end performance, or facilitate end-to-end channel estimation at the BS side via uplink UE pilot signal transmissions (see~\cite{RIS_Stc_mag} and references therein). In both cases, the BS finds, or computes, the optimum phase configurations for all RIS unit elements, which are conveyed to the RIS controller via the control channel. This information can be critical in certain applications, thus, needs to be secured in a way that can be supported by a very low RIS device.  

The feasibility of energy autonomicity of the previously described very low power RIS device has lately attracted various research efforts, being particularly beneficial in situations where the metasurface is deployed in remote or inaccessible locations, where it can be difficult or expensive to provide continuous power source supply~\cite{MEH}.
The main enabler for this autonomicity are the energy harvesting technologies, which target at harvesting energy from the incident radio waves at the metasurface. Such ambient-powered RIS devices capitalize on state-of-the-art metasurface-based radio-frequency energy harvesting solutions~\cite{amer_comprehensive_2020} 
to implement different harvest-and-reflect splitting mechanisms. The basic principle of metasurface energy harvesters is to capture radio waves in the ambient environment and convert them to guided waves, e.g., in a co-planar or  microstrip waveguide, or a coaxial cable, to empower the RIS controller, and consequently in our case study, the control communication of the RIS device with the BS~\cite{MEH}. 

As shown in Fig.~\ref{fig:RIS_control_channel}, data communication between the UE and the BS (solid red arrows) is already protected on the Packet Data Convergence Protocol (PDCP) and Radio Resource Control (RRC) layers for the user and control planes, respectively, which comprises integrity and confidentiality protection. As previously discussed, data communications between the UE and the BS is transparent to the RIS device, but the latter requires guidance from the BS for the optimal phase configurations of its response-tunable unit elements to serve the UE, thus, it requires a control configuration channel with the BS (dashed green arrow). It is also noted that, an RIS may be deployed in malicious ways to enable an optimized wireless link towards a legitimate system to facilitate successful decoding of the legitimate transmitted data signals~\cite{RIS_counteracting_2023,RIS_Attacks_2023}. Therefore, a secure configuration to the legitimate RIS is more crucial and it can prevent the malicious RIS involvement.

In the following, we present a lightweight security solution for confidentiality, integrity, and privacy protection of the control channel communication between the ambient-power RIS device and the BS in the system depicted in Fig.~\ref{fig:RIS_control_channel}.

\section{Secured Control of Ambient-Powered RISs} \label{Sec:Restrictions}
The security procedure for the control channel between a BS and an ambient-powered RIS device can be initiated by any of the two parties. The main idea is to use a secret parameter, only known to the RIS controller and the BS, as the basis for a simple key generation with a hash function. Hash functions are low computation intensive that can be carried out with the envisioned ambient-powered RIS use cases and scenarios~\cite{EURASIP_RIS_all}, considering the power and computational limitations of very low power RISs. To this end, the secret parameter can be the device IDentifier (ID) of the RIS, assuming that this ID is unique among all possible RIS device controllers within a smart wireless environment orchestrated by a single BS~\cite{RISE6G_COMMAG}. 
\begin{figure}[t!]
\centering
\includegraphics[width=\columnwidth]{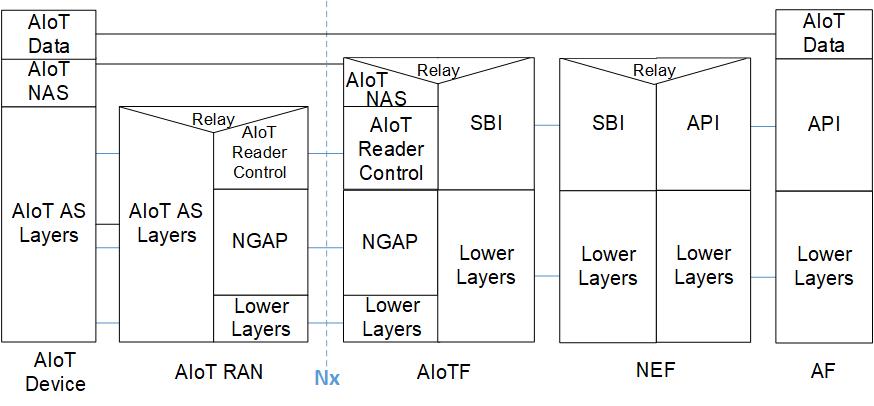}
\caption{The AIoT protocol stack as described in 3GPP TR 23.700-13~\cite{3gppTR2370013}.}
\label{fig:ambient_stack} 
\end{figure}
Extensions to smart wireless environments with multiple BSs and/or RISs shared (in time/frequency/space) among more than one BSs~\cite{EURASIP_RIS_all} are also feasible. The precondition is that some knowledge on this device ID needs to be available at the BS that controls it, which can be acquired during the RIS installation over the top, e.g., via Operations, Administration, and Maintenance (OAM). This implies that the RIS controller ID is considered as a shared knowledge which is used for authentication between this RIS device and its connected BS or group of BSs (for shared utilization). 

To enable security, the RIS controller ID should not be transferred in clear over the air interface. Instead, an unrelated default ID of the RIS device can be used for setting up the security context. A shared secret is used as an input to the key generation with a nonce, including also the length of the nonce (e.g., $0{\rm x}04$), using a preconfigured hashing algorithm~\cite{3gppTS33501} (e.g., SHA$256$, SHA$384$, and SHA$3$). The encryption key may be the truncated output of the hash to the most or the least significant bits (e.g., $64$, $128$, or $256$ bits), depending on the RIS controller capabilities. It is noted that the key length also depends on the envisioned encryption algorithm and the message length, as well as the processing and memory capabilities of the RIS controller. It is also assumed that the encryption and hashing algorithms, the key length, and the length of the temporary ID are preconfigured in the RIS controller and in the BS it will be connected to, otherwise, a capability exchange needs to take place. To facilitate the issue of addressing an RIS controller, the device might have assigned a default ID for an one-time use, at the time of the initial registration. To address the RIS identification for upcoming message exchanges, a new temporary ID is generated as remaining part of the truncated output hash of the encryption key generation. The BS and the RIS controller would generate the same temporary ID and the same encryption key in the same way. It is noted that the RIS controller may support advanced security algorithms and may use this default limited security settings to generate the security key for confidentiality, to perform an encrypted security capability exchange, and to update to more enhanced algorithms for the security setup.

The new AIoT protocol stack from the architecture study in~\cite{3gppTR2370013} is illustrated in Fig.~\ref{fig:ambient_stack}. As depicted, the AIoT device has a control protocol layer, the AIoT Non-Access Stratum (NAS) protocol, with the AIoT Function (AIoTF) being responsible for exchanging inventory messages as well as command requests and responses. With respect to our RIS-based case study in~Fig.~\ref{fig:RIS_control_channel}, the AIoT device is the ambient-powered RIS controller. Consequently, the AIoT Radio Access Network (RAN) is the BS/gNB that needs to instruct the AIoT device (i.e., the RIS controller). The AIoTF can be co-located with the AIoT RAN (i.e., the BS), since the latter needs to send the control messages directly to the RIS controller. This function is used for security context generation and renewal, and retrieves the shared secret from the Application Function (AF) via the Network Exposure Function (NEF). In that sense, the BS is enhanced with a Service-Based Interface (SBI).

\section{Proposed Lightweight Security Protocol} \label{Sec:Light_Solution}
\begin{figure}[t!]
\centering
\includegraphics[width=\columnwidth]{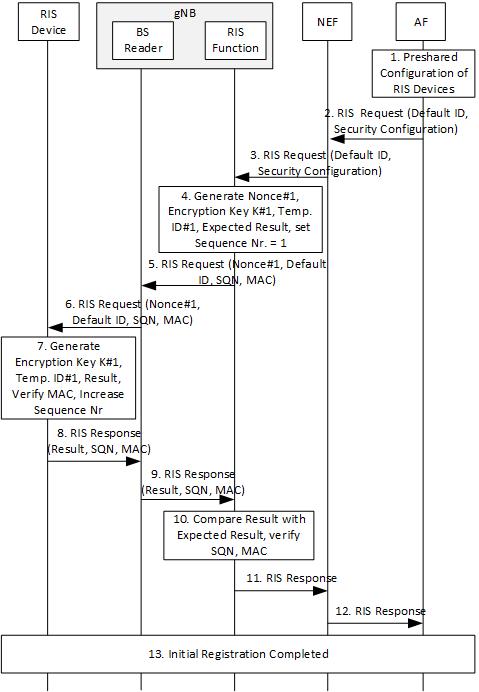}
\caption{The proposed lightweight security protocol for managing and operating an ambient-powered RIS device/controller through a BS that utilizes it to enable/boost its wireless connectivity with one or more UEs.}
\label{fig:Lightweight_Sec} 
\end{figure}
The proposed lightweight security solution assumes at least one shared device ID and a shared secret key between the ambient-powered RIS device and the BS that utilizes it to enable/boost its wireless connectivity with one or more UEs. The confidentiality key is firstly established, followed by the exchange of the integrity protection with a keyed Hash-based Message Authentication Code (HMAC) over the message to be protected. The functional split in the 3GPP architecture between the AIoTF and the BS reader is kept for the compliance with the 3GPP procedures, but both functions can be co-located in the BS/gNB. The AIoTF is renamed in our solution to RIS function, since the purpose is different: there are no inventory requests as well as no enabling/disabling command messages as in AIoT. Instead for RIS, the RIS function is required to send control messages to the RIS device (i.e., to its controller) without AF involvement.

The security setup solution proposed in this article is in line with the solution proposals in~\cite{3gppTR33713}, especially with solution $\#13$, one of the most comprehensive solutions in this technical report at the time of writing this paper. The proposal here is extended with some modifications to fit in our ambient-powered RIS framework. The overall procedure is described in Fig.~\ref{fig:Lightweight_Sec} and includes the following steps:
\begin{enumerate}[align=left]
     \item [\hspace{0.05cm} \textbf{Step 1:}]  The AF has a preshared configuration of the RIS devices, which includes a unique Default ID of the device and respective security parameters for deriving a security key, as well as Temporary IDs for ID privacy. The security parameters comprise a shared secret key, only known to the RIS Device and the AF. It is assumed that the AF possesses the geographical knowledge of which RIS Device will be served by the respective gNB/eNB.    
    \item [\hspace{0.05cm} \textbf{Step 2:}] The AF sends an RIS Request to the NEF with the Default ID together with the security parameters of the RIS Device.
    \item [\hspace{0.05cm} \textbf{Step 3:}] The NEF then forwards the RIS Request to the selected RIS Function.
    \item [\hspace{0.05cm} \textbf{Step 4:}] In the sequel, the RIS Function generates a Nonce and uses it to derive an Encryption Key together with a Temporary ID from the received security context from the NEF. The Encryption Key is used by this function to calculate an Expected Result. Then, the RIS Function sets the SeQuence Number (SQN) to “1” for the initial message. This number is then increased with every message received from the RIS Device and the RIS Function. The Encryption Key K and the Temporary~ID are computed as illustrated in Fig.~\ref{fig:ambient_2}a.
    \begin{figure*}
    \centering
    \includegraphics[width=\linewidth]{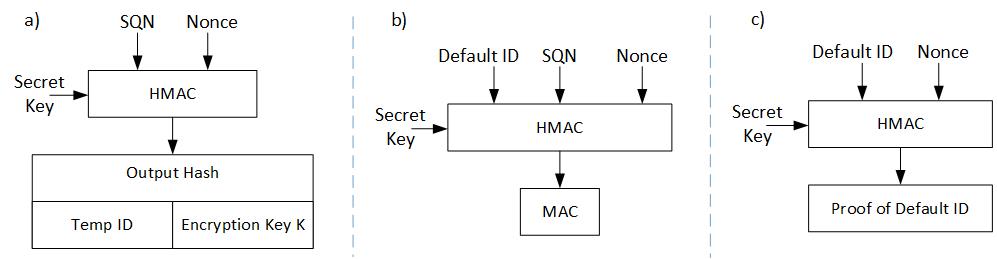}
    \caption{a) Derivation of a Temporary ID and an Encryption Key K; b) MAC generation; and c) Expected result/proof computation.}
    \label{fig:ambient_2} 
    \end{figure*}
    As shown in the figure, those elements in concatenation form the output of the HMAC function. It is noted that the split between the Temporary ID and the Encryption Key K may be equal, i.e., the lengths of the key and the ID can be the same. In particular, the Temporary ID may be formed from the most or least significant bits of the output hash, whereas the Encryption Key's bits are the remaining ones. Consequently, the Expected Result is a proof of the default identity, which is computed as depicted in Fig.~\ref{fig:ambient_2}c. 
    The RIS Function computes a Message Authentication Code (MAC) over the full message, i.e., using as inputs the Default ID, Nonce, and SQN, as shown in Fig.~\ref{fig:ambient_2}b.
    
    \item [\hspace{0.05cm} \textbf{Step 5:}] The RIS Function sends an RIS Request to the BS Reader. This request includes the Default ID and the Nonce, the Sequence Number (SQN), and the MAC.
    \item [\hspace{0.05cm} \textbf{Step 6:}]  The BS Reader sends the RIS Request to the controller of the RIS Device, which is tasked to listen to requests with the Default ID for initial gNB onboarding.
    \item [\hspace{0.05cm} \textbf{Step 7:}] The controller of the RIS Device verifies the MAC and the expected SQN. Then, it calculates the Encryption Key K and the Temporary ID for the next usage in a similar way as the RIS Function. The RIS controller also calculates the Result as a proof that it holds the security context, as shown in Fig.~\ref{fig:ambient_2}c. Afterwards, it increases SQN and protects the message with a MAC. For this, it deploys the proof as input, instead of the Nonce, to the HMAC function. The RIS controller is now ready to listen to the paging with the Temporary ID1, and expects the payload encrypted with the Encryption Key K1. 
    \begin{figure}[!t]
    \centering
    \includegraphics[width=\columnwidth]{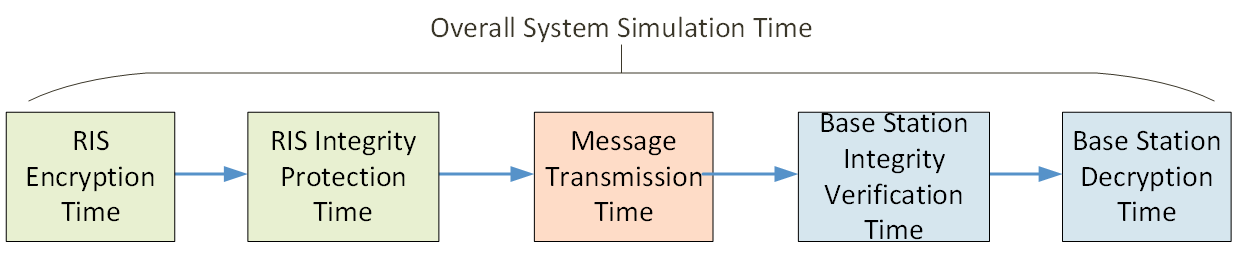}
    \caption{Transmission chain for simulating the system encryption/decryption time.}
    \label{fig:ambient_4} 
    \end{figure}
    \item [\hspace{0.05cm} \textbf{Step 8:}] The controller of the RIS Device sends an RIS Response to the BS Reader. This response includes the computed Result.
    \begin{figure*}
    \centering
    \includegraphics[width=\linewidth]{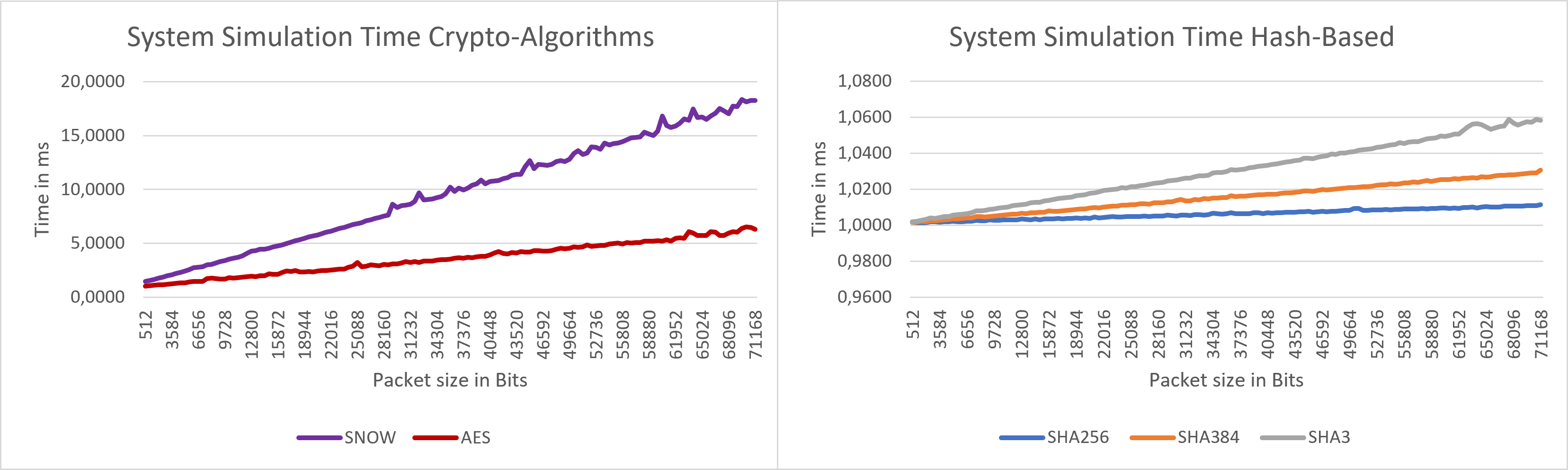}
    \caption{Comparison of the simulation time needed to implement a security protocol for the control channel between a BS and an ambient-powered RIS controller, considering USIM-based schemes (left) and the proposed lightweight scheme in Fig.~\ref{fig:ambient_4} with different hash functions (right).}
    \label{fig:ambient_5} 
    \end{figure*}
    \item [\hspace{0.05cm} \textbf{Step 9:}] Then, the BS Reader forwards the RIS Response to the RIS Function. 
    \item [\hspace{0.05cm} \textbf{Step 10:}] The RIS Function verifies the MAC and the expected SQN and compares the received result with the expected result and authenticates the RIS Device if both are identical. 
    \item [\hspace{0.05cm} \textbf{Step 11:}] In the sequel, the RIS Function sends an RIS Response to the NEF, indicating the success of the authentication.
    \item [\hspace{0.05cm} \textbf{Step 12:}] The NEF then forwards the RIS Response to the AF.
    \item [\hspace{0.05cm} \textbf{Step 13:}] The initial registration with the setup of the security association and authentication is completed. 
\end{enumerate}

All further requests from the AF following the completion of the initial registration via the proposed lightweight security protocol are  encrypted by the RIS Function and the RIS Device; the latter is addressed by only the Temporary ID. It is noted that the RIS Function may change the Encryption Key K and the Temporary ID by providing a new Nonce in a protected downlink request.

\section{Evaluation of Security Protocols for AIoT}\label{Sec:Outlook_Concl}
In this section, we compare the performance of the proposed lightweight security protocol in Section~\ref{Sec:Light_Solution}, designed for the control channel between a BS and an ambient-powered RIS device, with two algorithms for integrity protection used in the USIM in 5G for data protection. The encryption/decryption algorithm for the transmission chain described in Fig.~\ref{fig:ambient_4} was an Advanced Encryption Standard (AES) block cipher in all cases, and three different hash functions were compared, namely, SHA$256$, SHA$384$, and SHA$3$~\cite{3gppTS33501}. The integrity functions for the USIM-based approach were the AES and the word-oriented stream cipher SNOW~\cite{3gppTS33501}. The radio transmission time of each data packet was fixed at $1$ ms in our simulations. We have considered that the BS uses the same algorithm for integrity verification as the sender, as well as the same algorithm for decryption.  

Figure~\ref{fig:ambient_5} includes simulation time results for the considered versions of the proposed lightweight security protocol (left) and the USIM-based baseline ones (right). We have implemented all protocols on a laptop computer in C++, although implementation directly in hardware would be more efficient. Nevertheless, the implementation complexity in terms of execution time, and thus, energy consumption can be convincingly compared, since all simulations were performed on the same computer. It can be observed in the left subfigure that, due to the serial operation in the integrity algorithms AES and SNOW, their simulation time increases with the packet size in bits. This behavior is attributed to the complexity needed to compute the MAC for Integrity (MAC-I) of the packet on the sender side, and, on the receiver side, to verify the MAC-I of the sender. In the right subfigure, it is shown that, due to the fact that hashing with a secret parameter from the hash chain is used in the proposed protocol, the simulation time depends only slightly on the increasing packet size (smaller than $0.06$ ms), and remains substantially lower than the benchmark protocols, especially for large packet sizes. It can be also seen that there exist minor variations in the proposed protocol's simulation time (and consequently, computational complexity) due to the hash output size ($256$, $384$, and $512$ bits) and the Merkle-Damgård, or sponge, construction type~\cite{3gppTS33501}. Evidently, the protocol execution time is almost linear to the power consumption of the considered ambient-powered RIS; similarly will happen for any other AIoT device. 

\section{$6$G Outlook and Conclusion} \label{Sec:Outlook_Concl}
The integration of IoT, that refers to set of devices with restricted capabilities, and ambient computing, that refers to what happens when those devices are connected to the network, and possibly, operate collaboratively, or even learn certain attributes from each other, constitute key ingredients for the realization of the ambient intelligence paradigm envisioned for 5G-Advanced and 6G~\cite{3gppTR2370013}. Security is a critical factor for the effective functionality of individual AIoT devices as well as any ambient inter-IoT device communications (e.g., group communication)~\cite{IoT_AmI_2022}. In particular, secure communications in such an AIoT ecosystem contribute to accomplishing the common goal specific to the associated IoT applications: to provide service to any end-user(s) involved, e.g., in a living space or in a business operation. 

The recent concept of smart wireless environments enabled by the RIS technology, possibly including multi-functional metasurface devices (i.e., for programmable reflections, sensing, and even, wave-domain-based over-the-air computing)~\cite{RISoverview2023_all}, introduces novel devices within the network infrastructure, with recent advances elaborating on their ambient-powered versions~\cite{MEH,amer_comprehensive_2020}. State-of-the-art RIS devices with very low power consumption~\cite{etsiGR_RIS003}, which can be designed as AIoT devices, necessitate configuration from the BS that utilizes, and/or owns, them for optimal relaying of radio signals via a secure control communication channel. In addition, such a secure channel establishment with a legitimate RIS will prevent any malicious RIS involvement in the AIoT communication~\cite{RIS_counteracting_2023}. The lightweight security solution presented in this article can enable communication security for the proper functionality of AIoT devices, in particular, it has the potential to establish secure communications between an ambient-powered RIS controller and its managing BS. Future research can explore the potential to enable lightweight security solutions for control communications among ambient-powered multi-functional RIS devices, possibly shared among multiple BSs and deployed dynamically on demand~\cite{EURASIP_RIS_all}, as well as group communication scenarios. 

Data security issues in AIoT scenarios are also a concern, as ambient intelligence systems are connected to online networks that are vulnerable to cyber-attacks. Hackers may be able to access sensitive information, such as user credentials, financial data, or personal records, thus, an over-automation approach can also be a problem. To this end, there are several factors which influence the adoption of AIoT technologies, such as usability, technical feasibility, security, and confidence, as well as social and economic impacts, which have the potential to disrupt business models in several areas. However, ambient intelligence in RIS devices, which has been the core case study in this article, still needs to gain the confidence of mobile operators and the dependent partners, and this can be made possible predominantly with secure implementations of wide AIoT communication systems. This goal will require a lot of emphasis on accuracy, security, reliability, and privacy protection in future generations of wireless networks. 

\bibliographystyle{IEEEtran}
\bibliography{references}

\end{document}